\documentstyle[twocolumn,aps]{revtex}
\begin{document}
\draft
\title{Field induced long-range-ordering in an $S$=1 quasi-one-dimensional 
Heisenberg antiferromagnet}
\author{Z. Honda\cite{ZH00} and K. Katsumata}
\address{RIKEN (The Institute of Physical and Chemical Research),
Wako, Saitama 351-0198, Japan}
\author{Y. Nishiyama and I. Harada}
\address{Department of Physics, Faculty of Science, Okayama University, 
Okayama 700-8530, Japan}
\date{Received \today}
\maketitle
\begin{abstract}
Detailed results of heat capacity and magnetization measurements are reported on
a single crystal sample of the spin $S$=1 quasi-one-dimensional Heisenberg antiferromagnet
Ni(C$_{5}$H$_{14}$N$_{2}$)$_{2}$N$_{3}$(PF$_{6}$), abbreviated NDMAP.  
From these results, we constructed the magnetic field ($H$) versus temperature phase diagram
of this compound which exhibits the quantum disordered Haldane, field-induced long-range-ordered
(LRO) and thermally disordered paramagnetic phases.  The phase boundary curve separating the 
paramagnetic and LRO phases is anisotropic; the increase of the N\'{e}el temperature 
($T_{\rm N}$) with $H$ along the $a$ and $b$ axes is more rapid than that along the $c$ axis.
We calculated $T_{\rm N}$ as a function of $H$ by taking into account the inter chain coupling 
as the form of a mean-field.  The calculation of the staggered susceptibility of the $S$=1 
one-dimensional antiferromagnet with an easy-plane anisotropy shows quite different behavior in
the different field directions, resulting in the anisotropic phase boundary curve.  A good agreement
is obtained between theory and experiment using the exchange and anisotropy constants obtained
from the neutron scattering experiment. 
\end{abstract}

\section{Introduction}
Many lower-dimensional antiferromagnets show a singlet ground state and an energy gap
in their excitation spectrum originating from quantum many body effects.  The first 
indication of the quantum energy gap came out in 1983.  Haldane\cite{FDH83} conjectured 
that the excitation spectrum of one-dimensional (1D) Heisenberg antiferromagnets (HAFs) 
with integer spin quantum number ($S$) has an energy gap between the ground and first excited
states, while the corresponding system with half-odd integer $S$ has no energy gap.  This 
conjecture has been tested both theoretically\cite{IA89} and experimentally\cite{KK95}. 
It is now generally accepted that the quantum energy gap (Haldane gap) does exist in 1D 
HAFs with integer $S$.

In real materials, there always exists an interaction between spin chains.  We call this class
of materials a quasi-one-dimensional (Q1D) magnet.  Usually Q1D antiferromagnets exhibit
a long-range ordering (LRO) at finite temperature due to the inter-chain coupling ($J'$)\cite{LJ74}.
The compound Ni(C$_{2}$H$_{8}$N$_{2}$)$_{2}$NO$_{2}$(ClO$_{4}$), abbreviated NENP, 
is a typical example of $S$=1 Q1D HAFs\cite{AM82,JPR87} and does not show any indication
of LRO down to 300 $\mu$K\cite{OA92}, whereas the $S$=1 Q1D HAF CsNiCl$_{3}$ shows
LRO at $\sim$4.7 K\cite{DM77}.  The different behavior between the two compounds comes 
from the difference in the strength of $J'$.  How robust is the Haldane phase in a Q1D HAF against 
perturbation?  Sakai and Takahashi\cite{ST90} studied theoretically the ground state properties of 
an $S$=1 Q1D HAF with a single-ion anisotropy of the form $DS_{z}^{2}$.  They showed that
the Haldane disordered phase exists for $zJ'/J$$\leq$0.05 in a rather wide range of $D$ values, where
$z$ is the number of neighboring chains and $J$ is the intra-chain exchange interaction.

As was demonstrated experimentally\cite{KK89}, strong magnetic fields destroy the Haldane gap 
and the system recovers magnetism.  Then, we expect a magnetic ordering to occur in an $S$=1 Q1D 
HAF under high fields and at low temperatures.  The heat capacity measurement made on NENP in 
magnetic fields did not reveal any indication of LRO\cite{TK92}.  One of the reasons for this is that 
an external magnetic field applied along the chain axis of NENP induces a staggered field on the 
Ni$^{2+}$ sites because of the presence of two crystallographically inequivalent sites for 
Ni$^{2+}$\cite{MC91}.  This staggered field causes a small energy gap near the transition field from 
the disordered to the magnetized states\cite{MH94,SS94} and thus prevents the occurrence of LRO 
at low temperatures.

We have reported the observation of a field-induced LRO in the $S$=1 Q1D HAF
Ni(C$_{5}$H$_{14}$N$_{2}$)$_{2}$N$_{3}$(ClO$_{4}$), abbreviated NDMAZ\cite{ZH97},
and in Ni(C$_{5}$H$_{14}$N$_{2}$)$_{2}$N$_{3}$(PF$_{6}$), abbreviated 
NDMAP\cite{HAK98}.  Since there is only one site for Ni$^{2+}$ in NDMAZ and MAMAP,
these compounds are ideal for studying the field-induced LRO.  From a heat capacity ($C_{p}$)
measurement on a single crystal sample of NDMAZ, we have observed an anomaly at about 0.6 K and
12 T\cite{ZH97} which indicated that a magnetic ordering occurred there.  Because of limitations in 
our calorimeter, we were unable to follow how the position of the anomaly in $C_{p}$ changes with 
temperature ($T$) and magnetic field ($H$).  We then tried to synthesize the $S$=1 Q1D HAF, NDMAP, 
with a weaker $J$ than NDMAZ so that LRO is expected to be induced at a lower field.  We have 
measured $T$ and $H$ dependence of $C_{p}$ in a single crystal sample of NDMAP and constructed 
the $H$-$T$ phase diagram\cite{HAK98}.  We found an anisotropy in the phase boundary curve 
separating the paramagnetic and LRO phases, and we presented a qualitative interpretation for this 
anisotropy\cite{HAK98}.  Electron spin resonance (ESR) measurements made on a single crystal of 
NDMAP gave further evidence for the existence of the field-induced LRO phase and of the anisotropy 
in the phase boundary curve.\cite{HKH99}

In this paper we report detailed results of heat capacity and magnetization measurements made on
NDMAP.  We also report a theoretical analysis of the $H$-$T$ phase diagram.

The format of this paper is as follows.	In Sec.II we present the relevant background information
and details.  The experimental results are given in Sec.III.  In Sec.IV, a theoretical consideration is
given on the $H$-$T$ phase diagram.  The last section (Sec.V) is devoted to discussion and conclusions.

\section{Preliminary Details}
The compound Ni(C$_{5}$H$_{14}$N$_{2}$)$_{2}$N$_{3}$(PF$_{6}$)  (NDMAP) has the 
orthorhombic structure\cite{MM96} with the space group $Pnmn$ shown in Fig. 1.  The lattice 
parameters are, $a$= 18.046 $\AA$, $b$= 8.7050 $\AA$ and $c$= 6.139 $\AA$\cite{MM96}.
The structure consists of Ni(C$_{5}$H$_{14}$N$_{2}$)$_{2}$N$_{3}$ chains along the $c$ axis.
These chains are well separated from each other by PF$_{6}$ molecules.  All the Ni$^{2+}$ sites
in a chain are equivalent, $i.e.$, only one site exists for Ni$^{2+}$.

From the analysis of the magnetic susceptibility data, the following values are obtained\cite{HAK98}; 
$J/k_{B}$= 30.0 K, $D/J$= 0.3, $g_{\parallel}$= 2.10 and $g_{\perp}$= 2.17, where  $g_{\parallel}$
and $g_{\perp}$ are the $g$ values parallel and perpendicular to the chain $c$ axis, respectively.  Neutron 
inelastic scattering measurements were done on single crystals of deuterated NDMAP at $T$= 1.4 K
\cite{AZ00}.  From the analysis of the data, the values of exchange and anisotropy parameters are
determined to be, $J$= 2.28 meV (= 26.5 K), $J'_{x}$= 3.5$\times$10$^{-4}$ meV 
(= 4.1$\times$10$^{-3}$ K), $J'_{y}$= 1.8$\times$10$^{-3}$ meV (= 2.1$\times$10$^{-2}$ K) 
and $D$= 0.70 meV (= 8.1 K), where $J'_{x}$ and $J'_{y}$ are the inter-chain exchange interactions
along the $a$ and $b$ axes, respectively.

The single crystals of NDMAP used in this study were grown from an aqueous solution of NaN$_{3}$,
Ni(NO$_{3}$)$_{2}$$\cdot$6H$_{2}$O and 1,3-diamino-2,2-dimethylpropane.  After filtration, 
KPF$_{6}$ was added to the solution.  Well shaped blue single crystals up to
5mm$\times$5mm$\times$20mm were obtained after several weeks.  Fully deuterated single crystals
of NDMAP (NDMAP-d$_{28}$) were grown from a D$_{2}$O solution of the same ingredients 
except that a deuterated 1,3-diamino-2,2-dimethylpropane was used.

The single crystals thus obtained were checked by a four circle X-ray diffractometer.  We confirmed 
that the lattice parameters of our crystal are almost identical with those reported before\cite{MM96}.  
We measured $C_{p}$ of NDMAP and NDMAP-d$_{28}$ and found that the $H$-$T$ phase diagrams
of the two systems are essentially the same.  This means that the magnetic parameters in NDMAP and
NDMAP-d$_{28}$ are not different.

Heat capacity measurements were performed with a MagLab$^{\rm HC}$ microcalorimeter (Oxford Instruments, 
UK).  The temperature and magnetic field ranges accessible with this calorimeter are, 0.45 K$\leq$$T$$\leq$
200 K and 0$\leq$$H$$\leq$12 T.  Magnetization measurements were done with a MagLab$^{\rm VSM}$
vibrating-sample-magnetometer (Oxford Instruments, UK).  The temperature and magnetic field ranges
available with this magnetometer are, 1.5 K$\leq$$T$$\leq$300 K and 0$\leq$$H$$\leq$12 T.

\section{Experimental Results}
\subsection{Heat Capacity}
Figure 2 shows the temperature dependence of the heat capacity of NDMAP, including
the contribution of the lattice measured in zero field.  The data are well expressed by the following
equation,
\begin{equation}
C_{p} = a T + b T^{3}
\end{equation}	
in the temperature range between 2 and 5 K with $a$= 0.109 and $b$= 0.00653.  We use
hereafter this $b T^{3}$ term to subtract the contribution of lattice heat capacity as has been
done by many authors.

We show in Figs. 3(a)-(c) the temperature dependence of magnetic heat capacity ($C_{m}$) of NDMAP,
after subtracting the lattice heat capacity, in magnetic fields applied parallel to the $a$, $b$, and $c$ axes,
respectively.  In all field directions, we see an anomaly in $C_{m}$ at finite fields above a critical value. 
This anomaly signals that a magnetic ordering occurs there.

One of the advantages of our calorimeter is that a field dependent $C_{p}$ can be measured under a
constant temperature\cite{GS93}.  Strictly speaking, we need to change temperature to measure $C_{p}$.
However, the temperature increment necessary for the measurement is 0.5 - 1 \% of the temperature we
set so that temperature change during the measurement may be considered as small.  Figures 4(a)-(c) 
show such "field scan" data measured at several temperatures for the magnetic field directions parallel
to the $a$, $b$, and $c$ axes, respectively.  In addition to the sharp peak at the field denoted by 
$H_{\rm LRO}$, a broad feature is seen around the field named as $H_{\rm c}$.  Here, $H_{\rm LRO}$
means the position of $H$ at which the field-induced LRO occurs for a given $T$.

Combining all the information obtained from the heat capacity measurements, both of the "temperature
scan" and "field scan" procedures, we present the $H$-$T$ phase diagram of NDMAP in Fig. 5.  In addition
to the anisotropic phase boundary (curve (A)) separating the disordered and LRO phases\cite{HAK98}, 
we have another boundary (curve (B)) separating the Haldane and the disordered phases which is also
anisotropic.  The two curves (A) and (B) seem to merge at a finite $H$ when extrapolated to $T$= 0 K, 
for respective field directions.

\subsection{Magnetization}
Figures 6(a)-(c) show the temperature dependence of susceptibility (magnetization divided by applied
magnetic field, $M/H$) in NDMAP measured in magnetic fields applied parallel to the $a$, $b$, and $c$ 
axes, respectively.  The behavior of the susceptibility at $H$= 1 T is reminiscent of that taken at a much
lower field ($H$= 0.01 T)\cite{HAK98}; a broad peak around 35 K and a steep decrease in susceptibility
with decreasing temperature below about 20 K.  On increasing $H$, $M/H$ does not extrapolate to zero 
with $T$$\rightarrow$0.  This behavior of $M/H$ is similar to the one observed in an $S$=$\frac{1}{2}$
1D HAF\cite{LJ74} and indicates that a transition from the gapped to a gapless phase occurs at a higher field.
On increasing $H$ further, $M/H$ shows a minimum and an up turn at low temperatures.  The insets of 
Figs. 6(a)-(c) show the low temperature part of the data.  We see in the inset of Figs. 6(a) and (b) that
$M/H$ becomes almost temperature independent below a temperature whose value is field dependent.
The temperature independent susceptibility reminds us of the perpendicular susceptibility ($\chi_{\perp}$)
of an anisotropic antiferromagnet below the N\'{e}el temperature ($T_{\rm N}$).  We show below that
we are actually observing $\chi_{\perp}$ in this compound.  Because the sign of the single-ion anisotropy
term ($DS_{z}^{2}$) is positive in this compound, spins in the ordered phase are expected to lie in a plane
perpendicular to the $c$ axis (the quantization axis of the $D$ term is taken parallel to the $c$ axis).  The
anisotropy in the $c$ plane of the form $E(S_{x}^{2}-S_{y}^{2})$ is very small\cite{HKH99}.  Therefore,
when $H$ is applied along the $a$ or $b$ axes, spins point perpendicularly to $H$ in the $c$ plane keeping
an antiferromagnetic arrangement, thus giving $\chi_{\perp}$.  

We plot in Fig. 5 the transition points obtained from the $M/H$ data shown in Fig. 6(b).  These points
are defined as the temperatures where $d(M/H)/dT$ shows a minimum for a given $H$.  We see that the
transition points determined from the heat capacity and magnetization measurements agree well with
each other.

\section{ Theoretical Consideration on the Phase Diagram}
In this section, we try to reproduce the $H$-$T$ phase diagram observed in NDMAP, 
exhibiting an interesting behavior of $T_{\rm N}$ as a function of $H$: At low temperatures,
there occurs LRO only above a certain critical field and $T_{\rm N}$ shows an increase with increasing
fields, the rate of which depends on the direction of $H$.  We focus our attention especially on the 
physics behind the phase diagram.  To this end, we adopt the mean-field approximation 
for the interchain interaction\cite{JPA78}, which is known to work quite well except 
for fields in the vicinity of the critical field.  In the following, we use the energy unit $J$=1,
so that $t$, ${\bf h}$, $j'$ and $d$ are renormalized quantities of $T$, ${\bf H}$, $J'$ and $D$,
respectively.  According to the mean-field theory, the renormalized N\'{e}el temperature 
$t_{{\rm N}}({\bf h})$ as a function of the renormalized field ${\bf h}$ is given by the solution 
satisfying the following equation:
\begin{equation}
1/j'z = \chi_{{\rm st}}(t_{{\rm N}}({\bf h});{\bf h}),
\end{equation}
where $\chi_{{\rm st}}(t;{\bf h})$ is the staggered 
susceptibility for the {\em one-dimensional} magnetic system.  

Then, we calculate the staggered susceptibility of the magnetic chain by means of the quantum transfer
matrix method combined with the finite-temperature density matrix renormalization group.  As was
mentioned in the previous section, Ni$^{2+}$ spins in NDMAP has an easy-plane anisotropy, and thus 
the magnetic chain is well described by the following Hamiltonian:
\begin{equation}
{\cal H}= J [ \sum_n \{ {\bf S}_n\cdot {\bf S}_{n+1} + d (S_n^z)^2 \} \\
        - \sum_n g\mu_{\rm B}{\bf h}\cdot {\bf S}_{n}],
\end{equation}
where ${\bf S}_n$ represents the spin-$1$ operator at the $n$\/th site.  
Neglecting the small anisotropy in the $c$ plane, we consider the following two 
cases for the field ${\bf h}$ applied (i) along the $z$($c$)-axis 
(perpendicular to the easy plane) and (ii) along the $y$-axis (in the easy 
plane). 

It is noted here that the situation is quite different between the two 
cases, since the former field reserves the axial symmetry around 
the $c$-axis while the latter breaks it.  The nature of quantum fluctuations 
and hence the staggered susceptibility depends crucially on the symmetry 
of the system.  In the following, we consider the two cases separately.  

(i) {\it The field applied perpendicularly to the easy plane} (${\bf h}=
(0,0,h)$) \\
As was mentioned, below the critical field $h_{\rm c}$ the nonmagnetic 
Haldane phase called a quantum disordered phase is the ground state of 
the system, while above it the so called Tomonaga-Luttinger liquid state 
becomes the ground state.  Although the former has an excitation gap to 
the triplet state, the latter has a gapless excitation spectrum and hence 
is critical.  This criticality of the Tomonaga-Luttinger liquid is 
characterized by the critical exponent $\eta$ defined by the divergence of 
the staggered susceptibility at zero temperature\cite{ST91}:
\begin{equation}
\chi_{{\rm st}}(t;h) = A t^{-(2-\eta)}, 
\end{equation}
where $A$ is a constant, which scarcely depends on ${\bf h}$ in our 
calculation.  Note that in the classical system $\eta=0$.  We may safely use
Eq. (4) at low temperatures well below the temperature at which the susceptibility
is maximum ($\sim$35 K).  In Fig. 7, $\eta$'s, estimated from the numerical calculations
for the temperature region $0.1<t<1.0$, are shown by the solid circles, each of which
has an error bar of $\pm 0.05$.  The data have been shifted in fields so that the 
theoretical critical field coincides with the one observed.  The solid curve
represents the phenomenological relation between $\eta$ and $h$: 
$\eta =0.3 {\rm exp}\{-\beta(h-h_{\rm c})\}+0.2$ with 
$\beta=0.5$ and $h_{\rm c}=0.2$.  

Now, we reproduce the phase diagram, using Eqs. (2) and (4) with $J=26.5 $K\cite{AZ00}.
In Fig. 8 we show the theoretical result by the solid curve and the experimental data by
the solid circles.  Here, we adjusted the theoretical curve to reproduce the experimental point
$T_{\rm N}=0.92$ K at 11 T.   From Figs. 7 and 8, we see that the increase of $T_{\rm N}$
with $H$ is a consequence of the decrease in $\eta$ with $H$.  Remembering that $\eta$ 
measures the degree of quantum fluctuations, we can say that the quantum fluctuation out 
of the easy plane is reduced by $H$ so that the N\'{e}el state becomes more stable.  In contrast to
this quantum system, the field dependence of the constant $A$ is an only source of the field
dependence of $T_{\rm N}$ in the classical system, being very mild.  We show in Fig. 7 $\eta$
estimated from the phase diagram (Fig. 5) by the open circles, which follows also the
phenomenological relation but $\beta=1.0$.  Considering the large error bars in $\eta$ of
our estimations, we do not think the discrepancy in $\beta$ so seriously.  Observation of
the field dependence of $\eta$ by other methods is desired.

In stronger fields, spins cant in the field direction and hence $A$ decreases 
seriously. Thus, the phase boundary curve closes at the upper critical field, 
where the magnetic moment saturates.  From the phase diagram, we estimate 
the interchain coupling $j'z$ to be $1.2\times10^{-3}$, using the value 
$A=2.0$.  This value is favorably compared with that obtained from the neutron
inelastic scattering experiments ( 2$j'_{x}$+2$j'_{y}$=$1.9\times10^{-3})$\cite{AZ00}.

(ii) {\it The field applied in the easy plane} (${\bf h}=(0,h,0)$) \\
In this case, above $h_{\rm c}$, the ground state has the N\'{e}el order 
and a gap opens again in the excitation spectrum because of the symmetry
breaking field.  This situation is quite different from the former case.  The
field dependence of $T_{\rm N}$ reminds us of the {\em soliton scenario} 
in the classical system\cite{HSH81}.  Remembering the form of the staggered
susceptibility in the classical system, we postulate the following in this case:
\begin{equation}
\chi_{{\rm st}}(t;h) = (B/t^2) {\rm exp}\{\alpha (h-h_{\rm c})/t\}, 
\end{equation}
where $B$ is a constant.  This form is confirmed by our numerical 
calculations, in the temperature region $0.1<t<1.0$, with the coefficient 
$\alpha$, a little less than unity.  Although $\alpha$ may represent 
quantum effects in the formation energy of soliton, we assume, for 
simplicity, the classical value 1 for it.  

Now, we again reproduce the phase diagram for this case using Eqs. 
(2) and (5).  The theoretical curves are adjusted as before using the 
experimental point $T_{\rm N}=2.2$ K at 12 T and the critical field value
$H_{\rm CF}=5.7$ T for ${\bf h}$ parallel to the $a$-axis, and $T_{\rm N}=2.7$ K
at 12 T and $H_{\rm CF}=5.4$ T for ${\bf h}$ parallel to the $b$-axis, respectively.
The different values of $H_{\rm CF}$ are due to an in-plane anisotropy, 
being neglected in this paper.  The phase boundary curves are shown by the solid 
curve for ${\bf h}$ parallel to the $a$-axis and by the dotted curve 
for ${\bf h}$ parallel to the $b$-axis with the corresponding 
experimental points, respectively, by the solid and the open circles 
in Fig. 9.  The agreement between theory and experiment is satisfactory.
We mention that the soliton scenario still works in our quantum system: 
The symmetry breaking field yields the uniaxial symmetry and hence the soliton
is a dominant source for the fluctuations in this system.  Since the soliton 
formation energy, i.e. the gap, increases with $h-h_{\rm c}$ and hence the 
staggered susceptibility increases exponentially at low temperatures, $T_{\rm N}$
shows a rapid increase with $H$, the rate of which is marked contrast 
with the former case.  Although the soliton scenario is effective in our case, more 
sophisticated study is required to establish further a quantum analogue of the soliton
in classical spin chains.

The phase boundary curve closes also at the upper critical field as in the former case.
We estimate the interchain coupling $j'z$ to be $1.4\times10^{-3}$, using the value 
$B=0.6$.  The value $j'$ estimated again agrees with the value observed\cite{AZ00}.

\section{Discussion and Conclusions}
In Sec. III, we have presented detailed results of heat capacity and magnetization measurements
on NDMAP from which we have constructed the $H$-$T$ phase diagram shown in Fig. 5.  We have
been successful in explaining theoretically the phase boundary curve separating the paramagnetic 
and LRO phases in Sec. IV using the values of intra-chain exchange interaction and anisotropy
constants determined from the neutron inelastic scattering measurements\cite{AZ00}.  The inter-chain
exchange interactions estimated theoretically are close to those obtained from the neutron experiment\cite{AZ00}.

We discuss the lower field boundary separating the Haldane and paramagnetic phases (curve (B) in
Fig. 5).  We argue below that the anomaly in $C_{m}$ observed along this curve is due to
the field dependence of the first excited triplet\cite{KK89,OG93}.  We analyzed the low 
temperature part of the magnetic heat capacity data using a two-level system model, with a singlet ground
state and the lowest state of the excited triplet with an energy difference ($\Delta_{-}(H)$), which gives a Schottky
type anomaly.  We show in Fig. 10, $\Delta_{-}(H_{x})$, $\Delta_{-}(H_{y})$ and $\Delta_{-}(H_{z})$
thus obtained for the field directions parallel to the $a(x)$, $b(y)$ and $c(z)$ axes, respectively.  The
solid curves in Fig. 10 represent the theoretical energy level as a function of $H$\cite{OG93}.  Here,
we used the value determined from the neutron inelastic scattering experiment\cite{AZ00} for the energy
gap in respective field directions at $H$$\rightarrow$0.  We see in this figure that the agreement between
theory and experiment is satisfactory.

Finally, we discuss the temperature dependence of $M/H$ (Figs. 6(a)-(c)).  We calculated $M/H$ as
a function of $T$ using the quantum transfer matrix method with a density-matrix renormalization 
group technique.  We compare theory and experiment in Fig. 11(a) for selected values of $H$ parallel
to the $a$ axis.  Here, we used $J/k_{\rm B}$=26.5 K obtained from the neutron inelastic scattering
study\cite{AZ00} and $g$=2.14 determined from the ESR measurement\cite{HK00}.  We see a good
agreement between theory and experiment without any adjustable parameters.  We have obtained the 
value $J/k_{\rm B}$=30.0 K from a fitting of the theory with the susceptibility data at high temperature 
range above about 40 K\cite{HAK98}.  Because the lattice parameters change with temperature\cite{MM96},
it is not surprizing if the exchange interaction constant determined at low temperatures is different from
that at high temperatures.  Figure 11(b) shows the case when $H$ is applied along the $c$ axis.  Since no
ESR data are available along this direction, we assumed the value 2.05 for $g$.  The agreement between
theory and experiment is not as good as in Fig. 11(a).  Further study is necessary to clarify this point.

In conclusion, we have reported detailed results of heat capacity and magnetization measurements on
a single crystal sample of the $S$=1 Q1D HAF, NDMAP.  From these results, we constructed the
$H$-$T$ phase diagram which exhibits the quantum disordered Haldane, field-induced LRO and
thermally disordered paramagnetic phases.  The phase boundary curve separating the paramagnetic
and LRO phases is anisotropic; the increase of $T_{\rm N}$ with $H$ along the $a$ and $b$ axes
is more rapid than that along the $c$ axis.  We calculated $T_{\rm N}$ as a function of $H$
by taking into account the inter chain coupling as the form of a mean-field.  We first evaluated
numerically the staggered susceptibility of the $S$=1 1D antiferromagnet for $H$ applied
perpendicularly to the easy plane, which shows, at low temperatures, a typical divergence of the
Tomonaga-Luttinger liquid with the critical exponent $\eta$.  Then, we got a satisfactory agreement
with the experimental results using the exchange and anisotropy constants obtained from the neutron
scattering experiment\cite{AZ00}.  It is interesting to note that the transition temperature $T_{\rm N}$
is governed by the critical exponent $\eta$ of the Tomonaga-Luttinger liquid.  On the other hand, for $H$
applied in the easy plane, we invoked the soliton scenario and got again a satisfactory agreement
with experiment.

\section*{Acknowledgements}
This work was partially supported by a Grant-in-Aid for Scientific Research from the Japanese 
Ministry of Education, Science, Sports and Culture.  Z. H. was supported by the Research Fellowships 
of the Japan Society for the Promotion of Science for Young Scientists.  The computation in this work
has been done using the facilities of the Supercomputer Center, ISSP, University of Tokyo.

\newpage
\begin{figure}
\caption{The crystal structure of  Ni(C$_{5}$H$_{14}$N$_{2}$)$_{2}$N$_{3}$(PF$_{6}$)
abbreviated NDMAP.}
\label{fig1}
\end{figure}

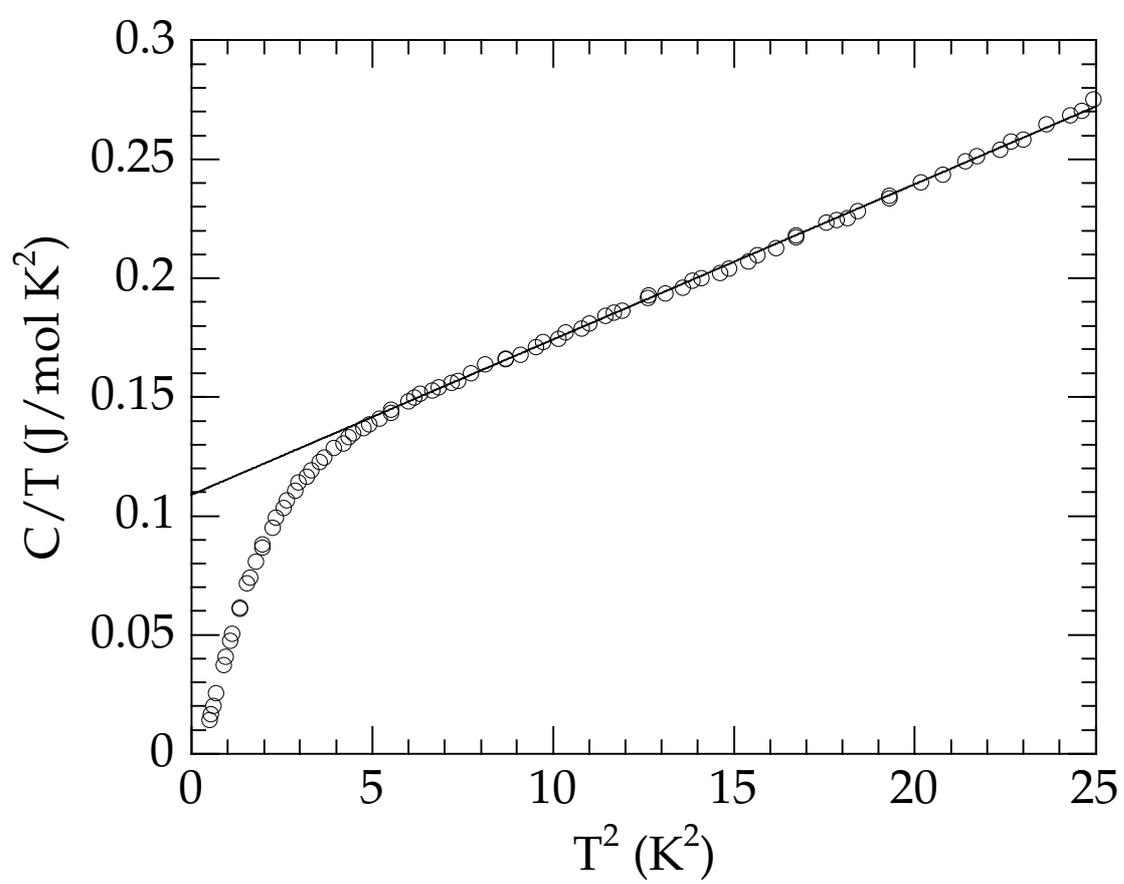
\begin{figure}
\caption{The temperature dependence of the heat capacity of NDMAP in zero external
magnetic field.}
\label{fig2}
\end{figure}

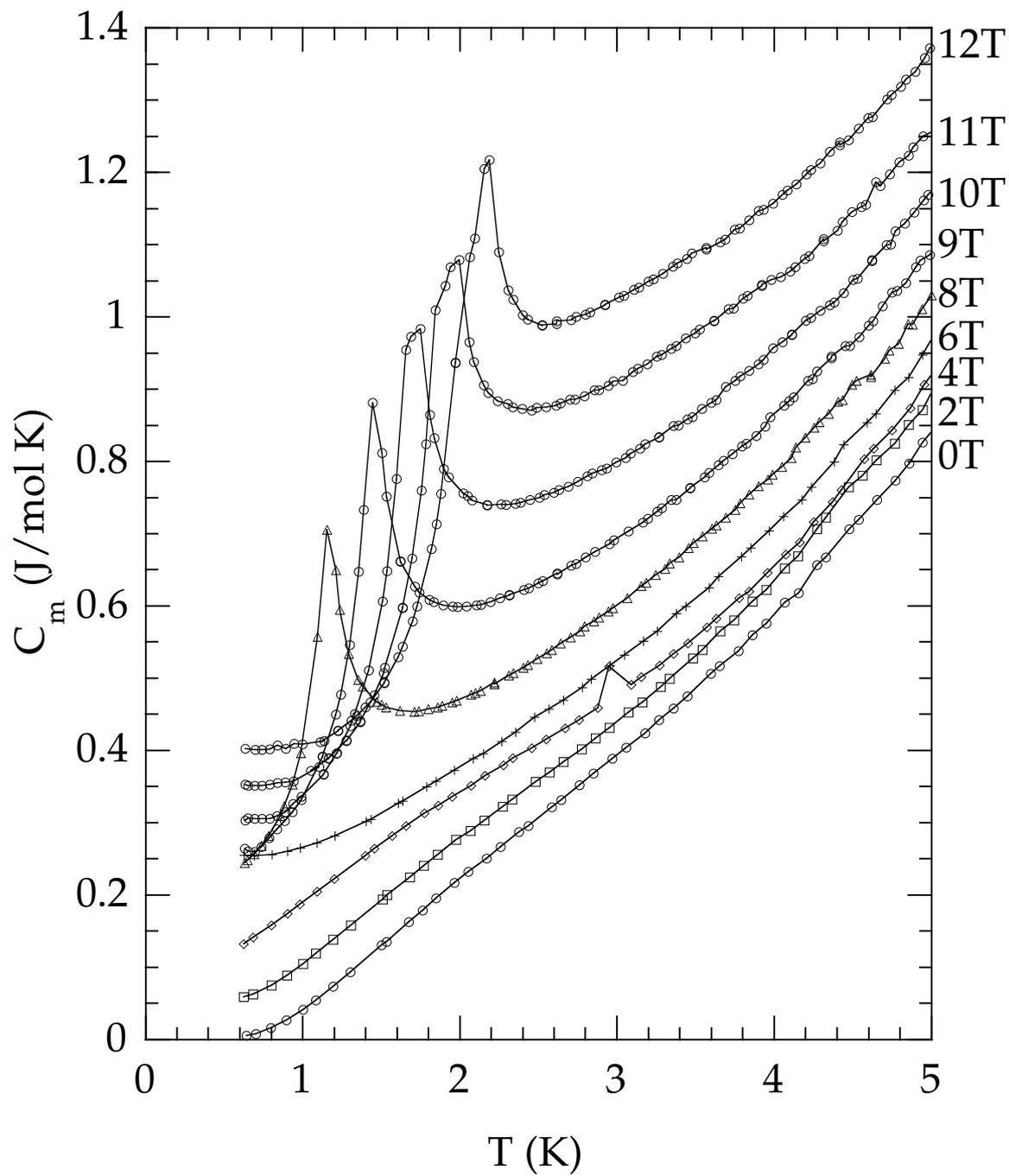
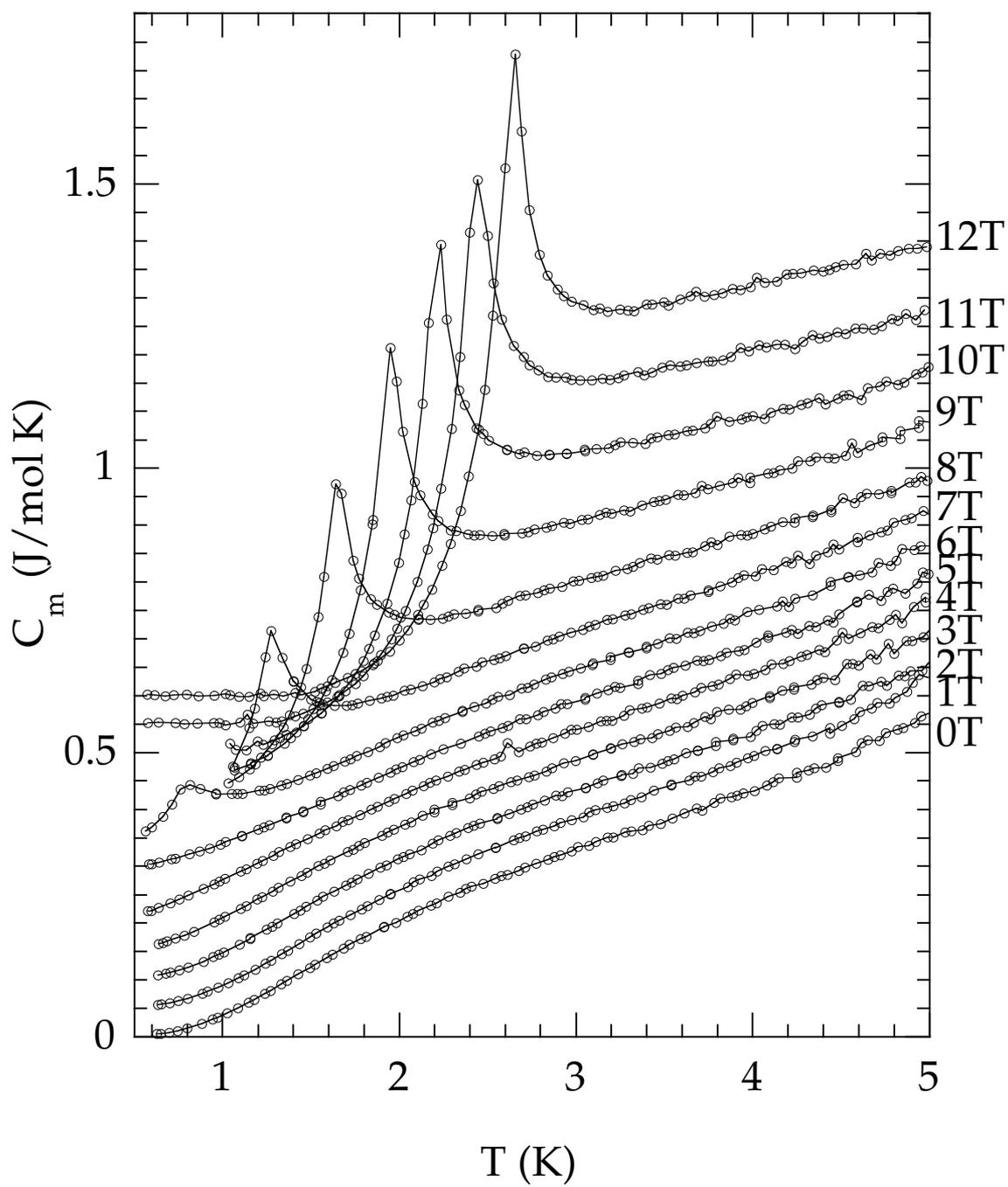
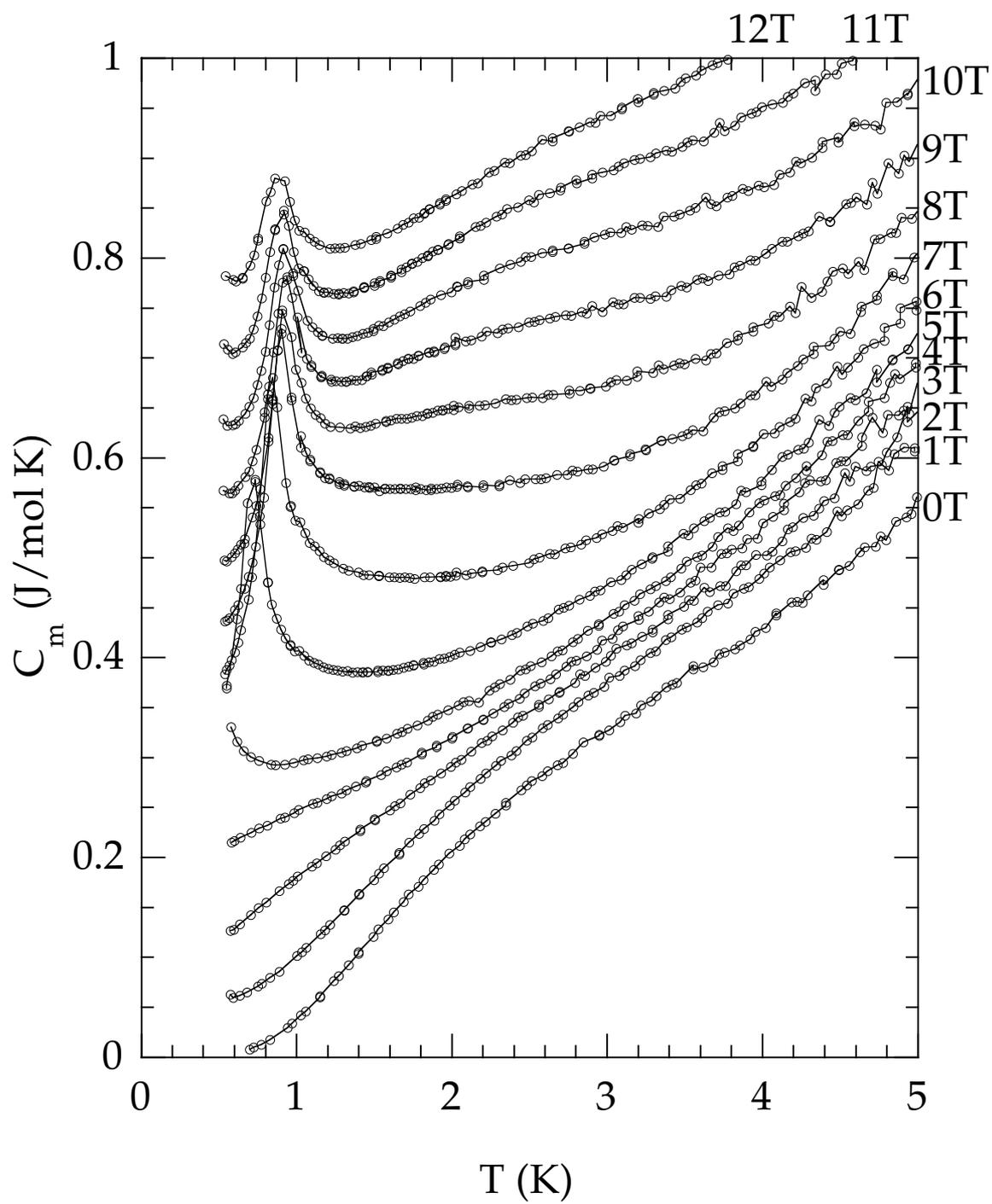
\begin{figure}
\caption{The temperature dependence of the magnetic heat capacity of NDMAP
measured at the designated fields applied along the (a) $a$, (b) $b$ and (c) $c$ axes, respectively.}
\label{fig3}
\end{figure}

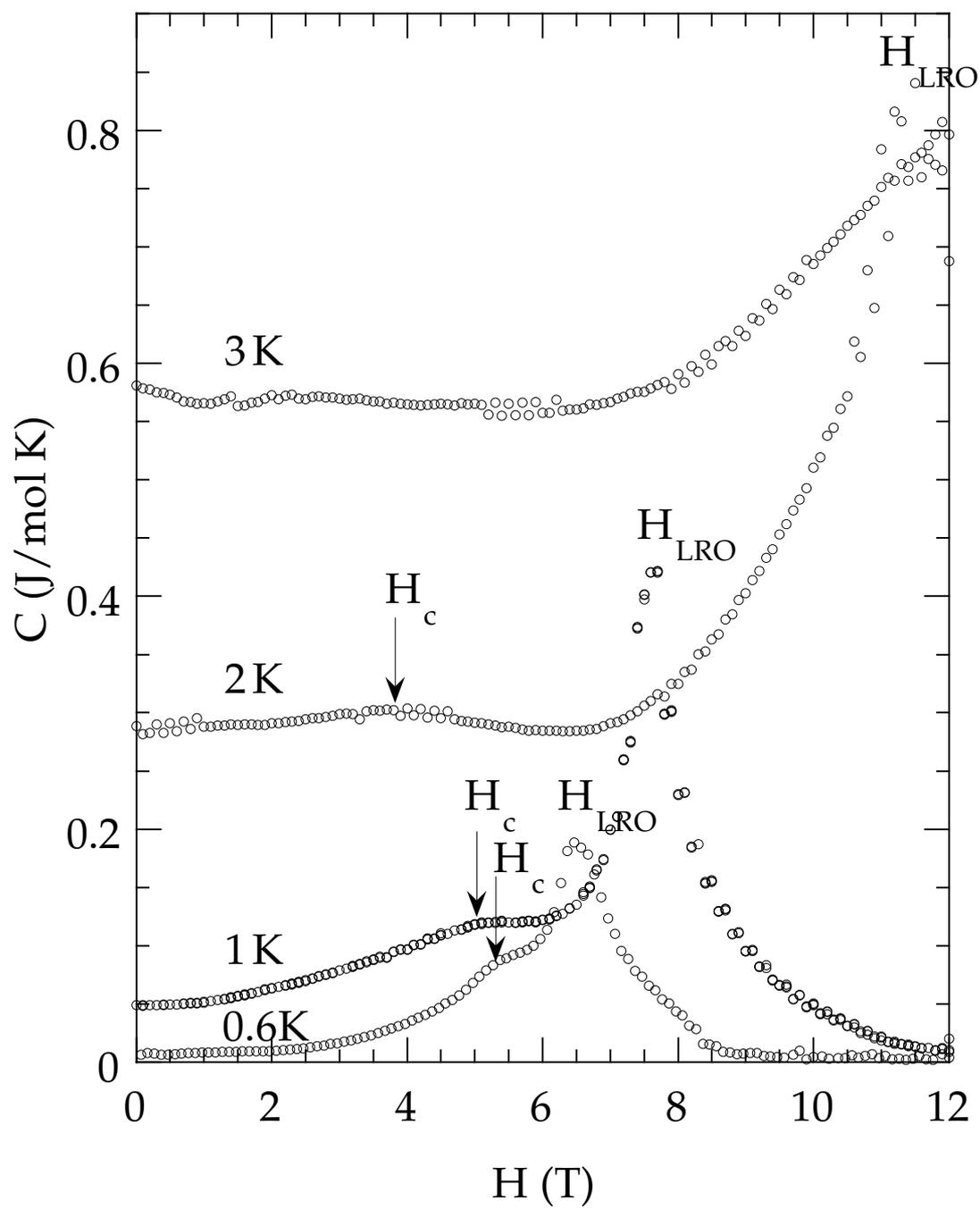
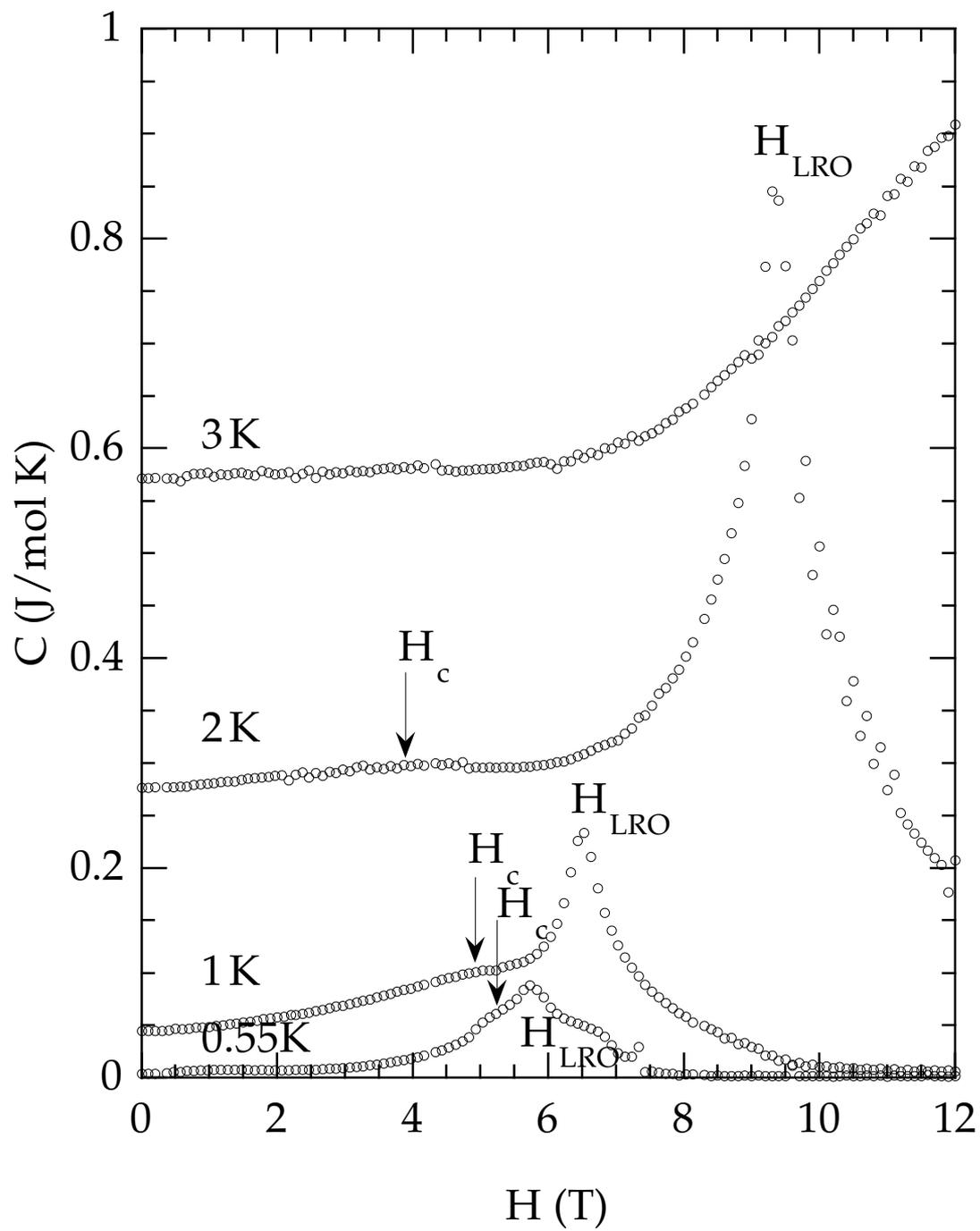
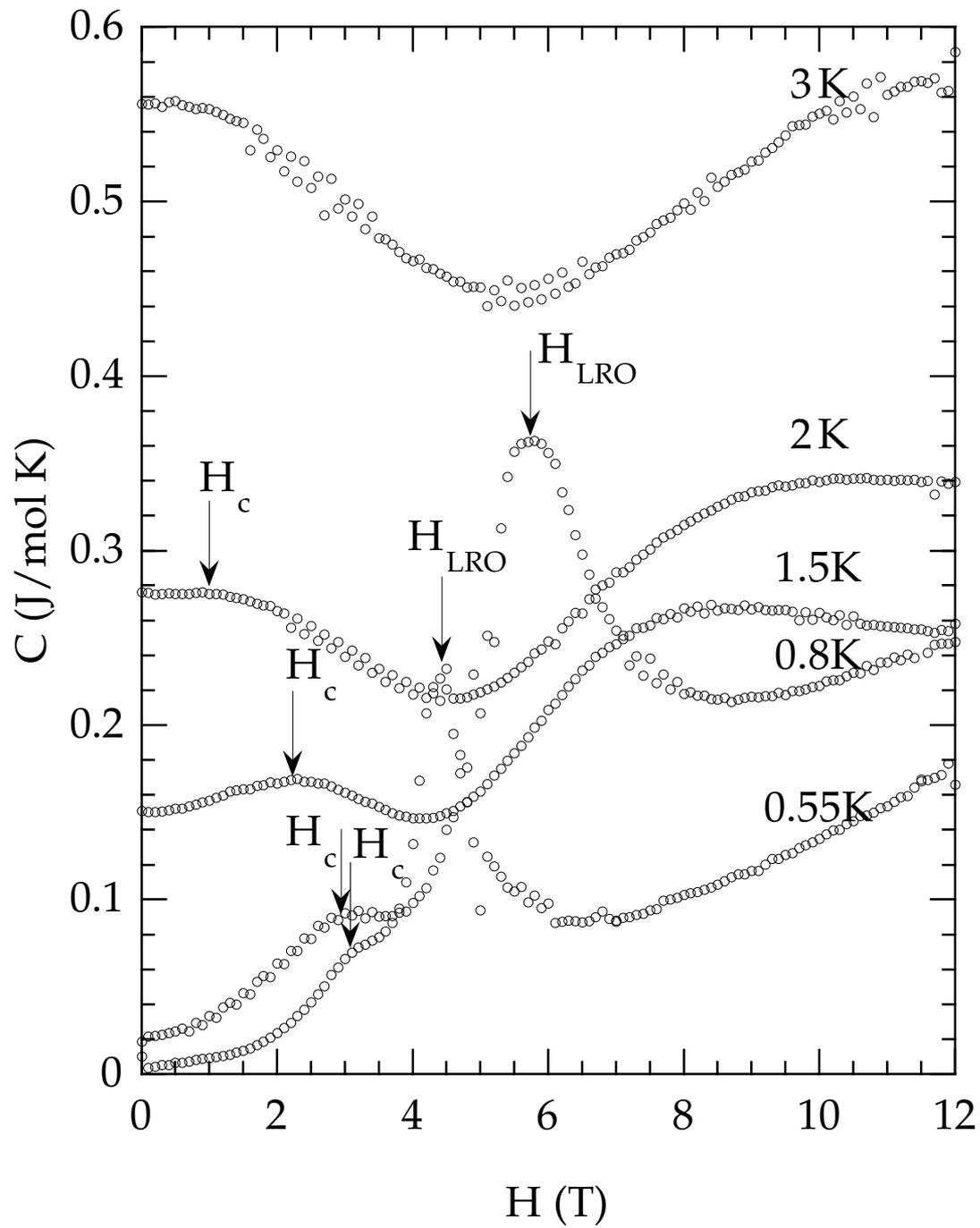
\begin{figure}
\caption{The magnetic field dependence of the heat capacity of NDMAP measured at
several temperatures.  The external magnetic field is applied parallel to the (a) $a$, (b) $b$ and
(c) $c$ axes, respectively.}
\label{fig4}
\end{figure}

\begin{figure}
\caption{The temperature vs. magnetic field phase diagram of NDMAP determined from the heat 
capacity measurements ($\bigtriangleup$, $\bigcirc$, $\Diamond$).  Also shown in this figure
are the transition points obtained from the magnetization measurements ($\times$).  LRO: 
long-range-ordered phase, H: quantum disordered Haldane phase. P: thermally disordered paramagnetic
phase.  Lines are a guide to eyes.}
\label{fig5}
\end{figure}

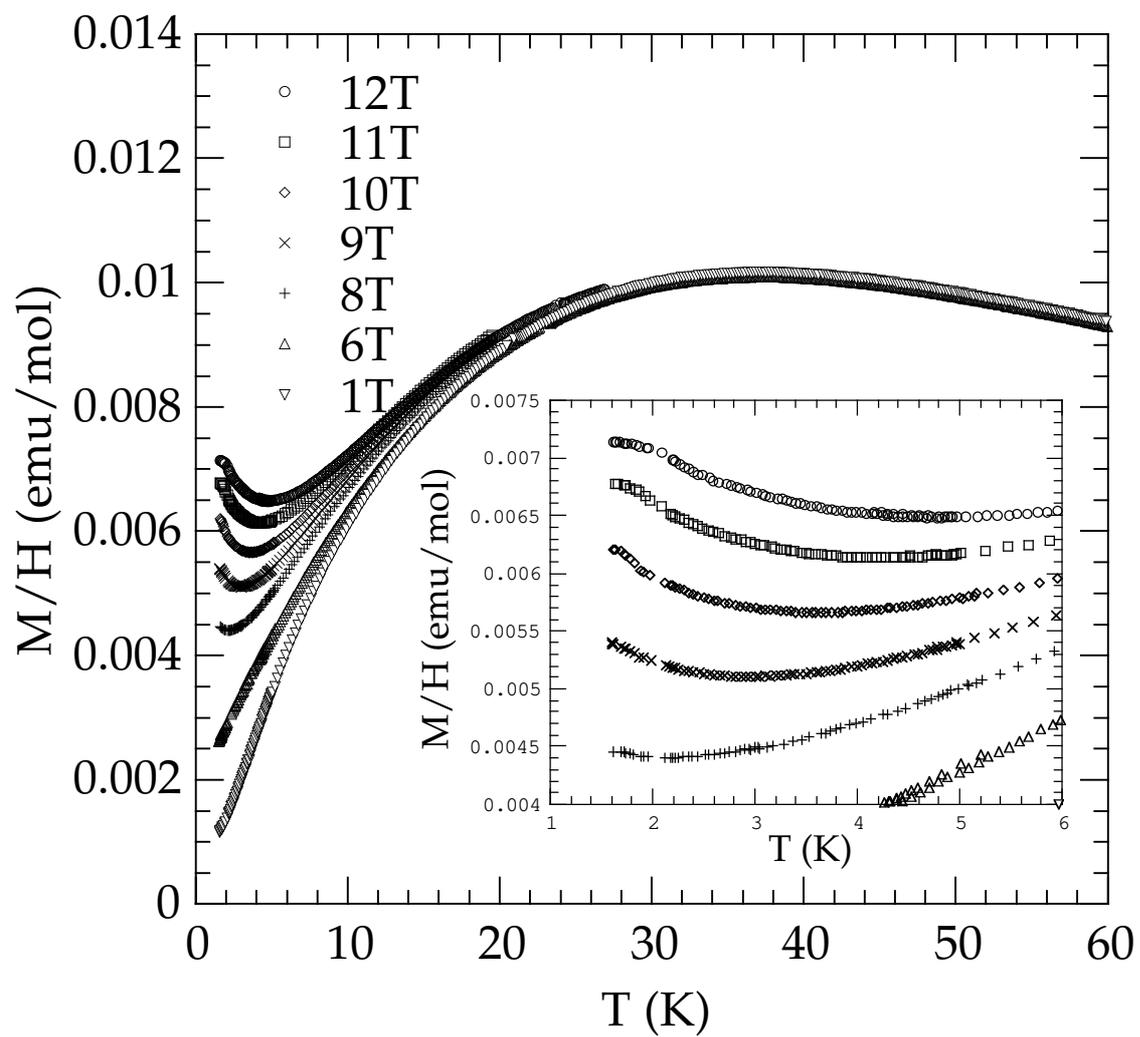
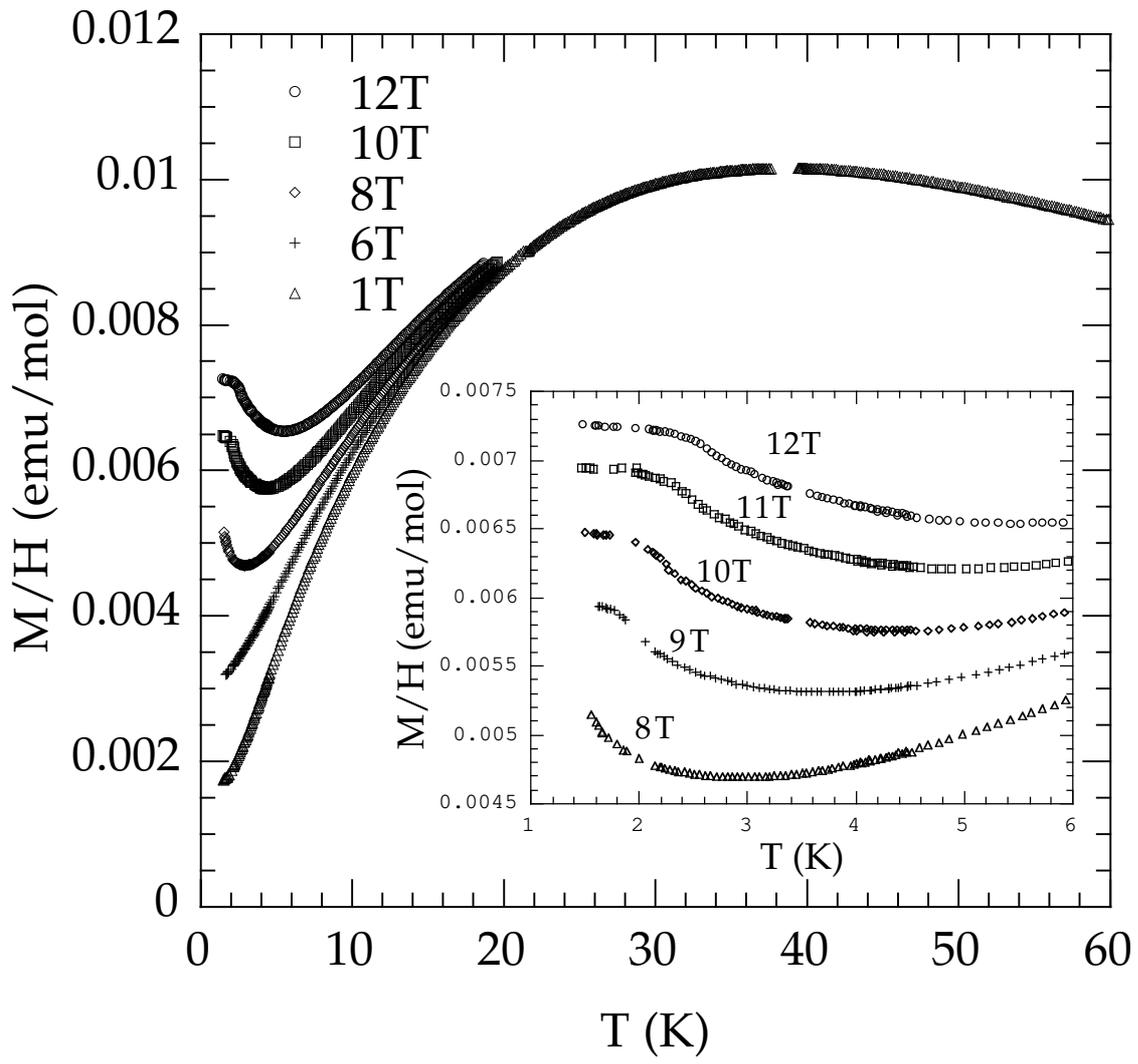
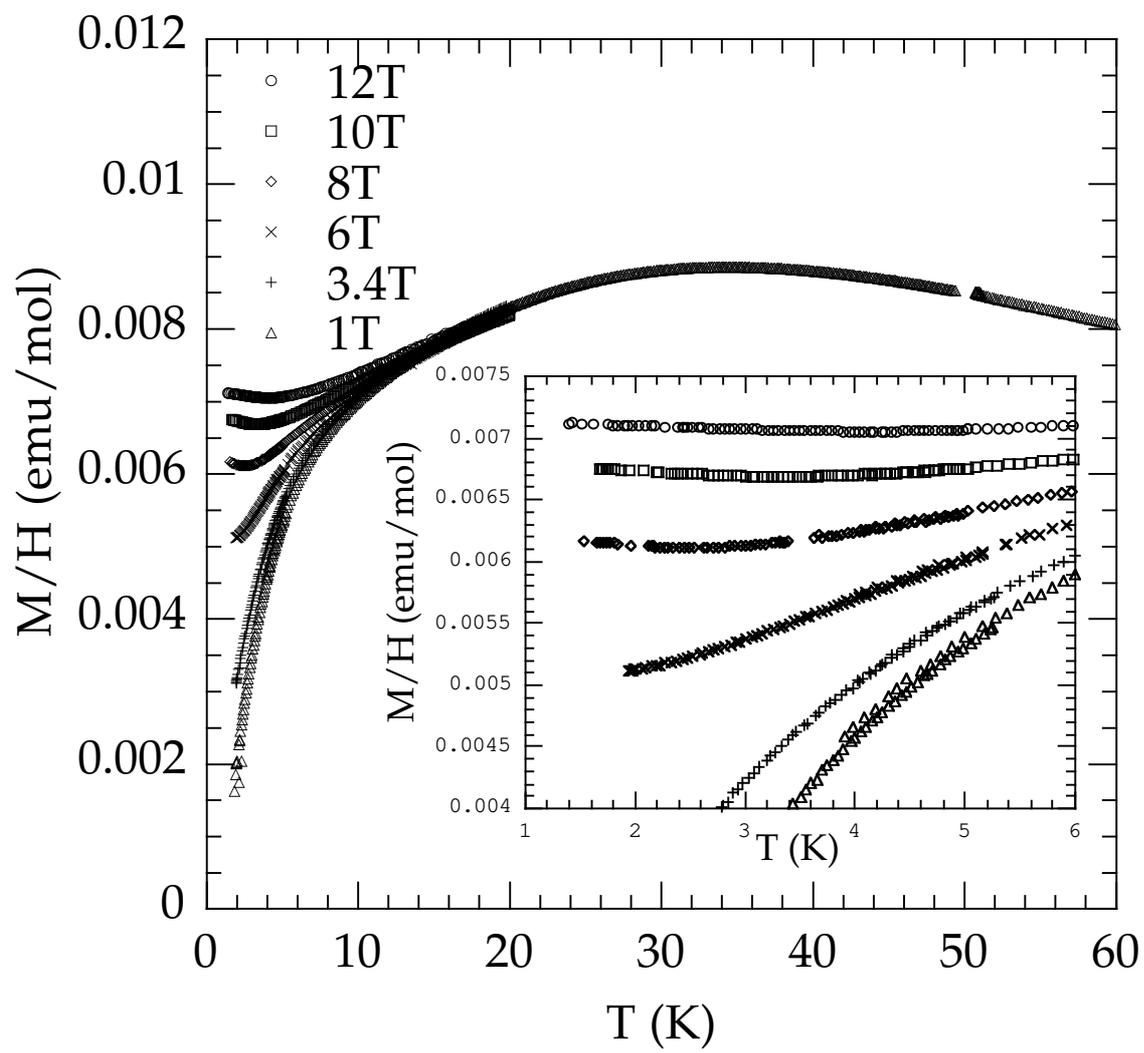
\begin{figure}
\caption{The temperature dependence of the susceptibility ($M/H$) in NDMAP measured at the
designated magnetic fields applied parallel to the (a) $a$, (b) $b$ and (c) $c$ axes, respectively.
The insets show the low temperature part of the data.}
\label{fig6}
\end{figure}

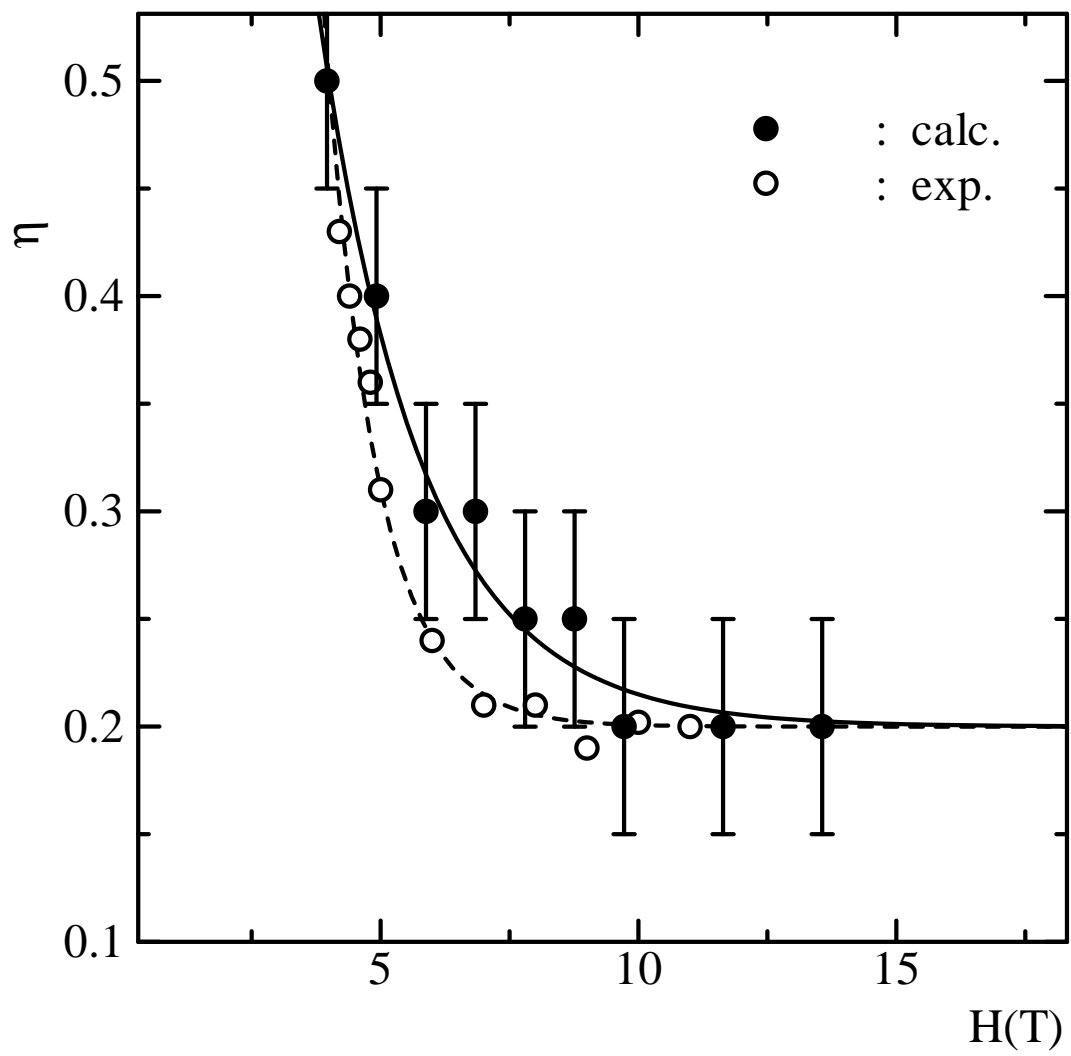
\begin{figure}
\caption{Field dependence of the critical exponent $\eta$ characterizing the 
Tomonaga-Luttinger liquid.   The solid circles denote $\eta$'s calculated 
for $d=0.3$ and the open circles are $\eta$'s estimated from the experimental 
phase diagram.  The solid and the dotted curves represent the phenomenological 
relations reproducing the calculated and the estimated results, respectively. 
(See text.)}
\label{fig7}
\end{figure}

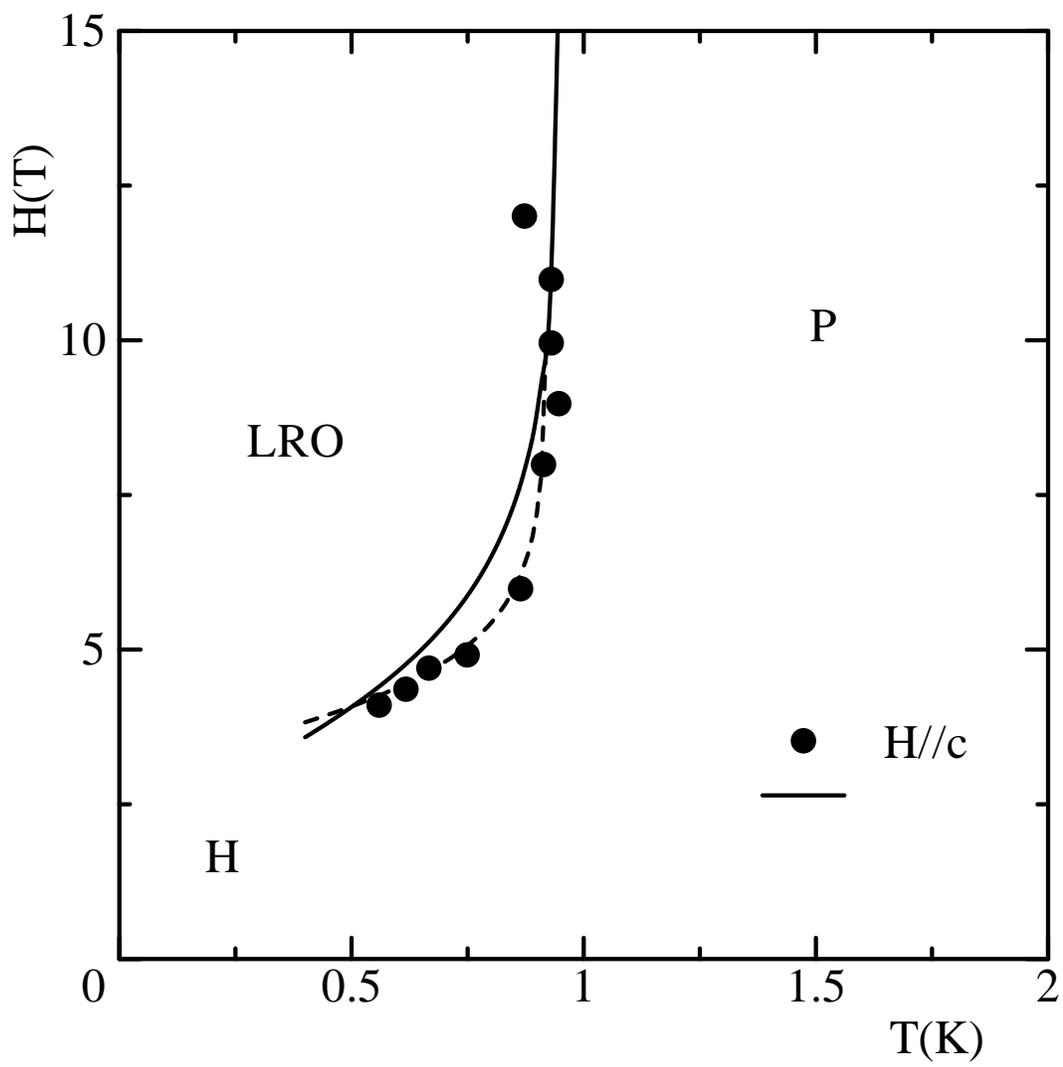
\begin{figure}
\caption{The $H$-$T$ phase diagram for the field applied perpendicularly to the easy 
plane ($H$$\parallel$$c$). The solid circles represent the experimental result while 
the solid curve represents the calculated result.}
\label{fig8}
\end{figure}

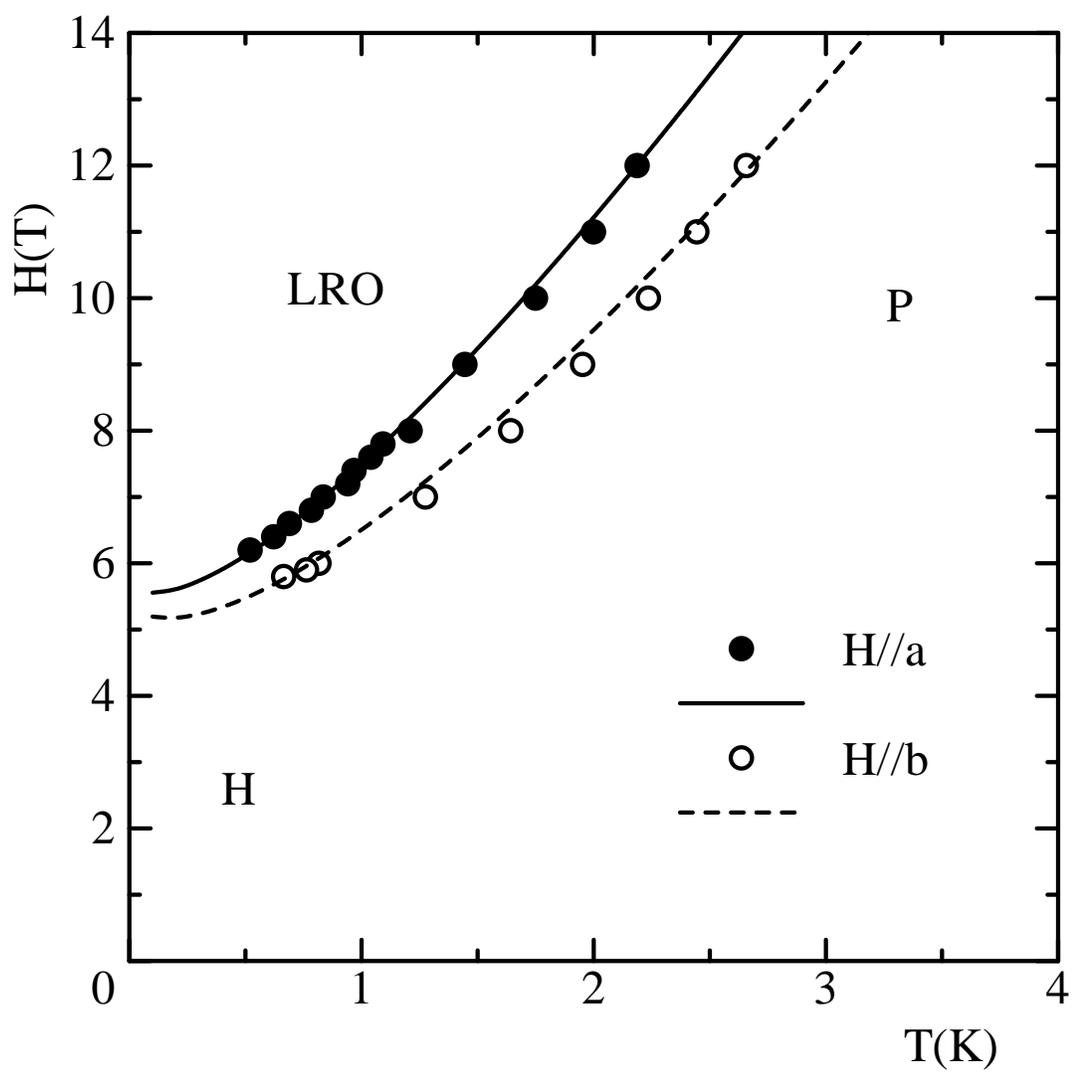
\begin{figure}
\caption{The $H$-$T$ phase diagram for the field applied in the easy plane. 
The solid  and open circles represent the experimental results, 
respectively, for the field parallel to the $a$ axis and for the field 
parallel to the $b$ axis, while the solid and the dotted curves denote 
the corresponding theoretical results.}
\label{fig9}
\end{figure}

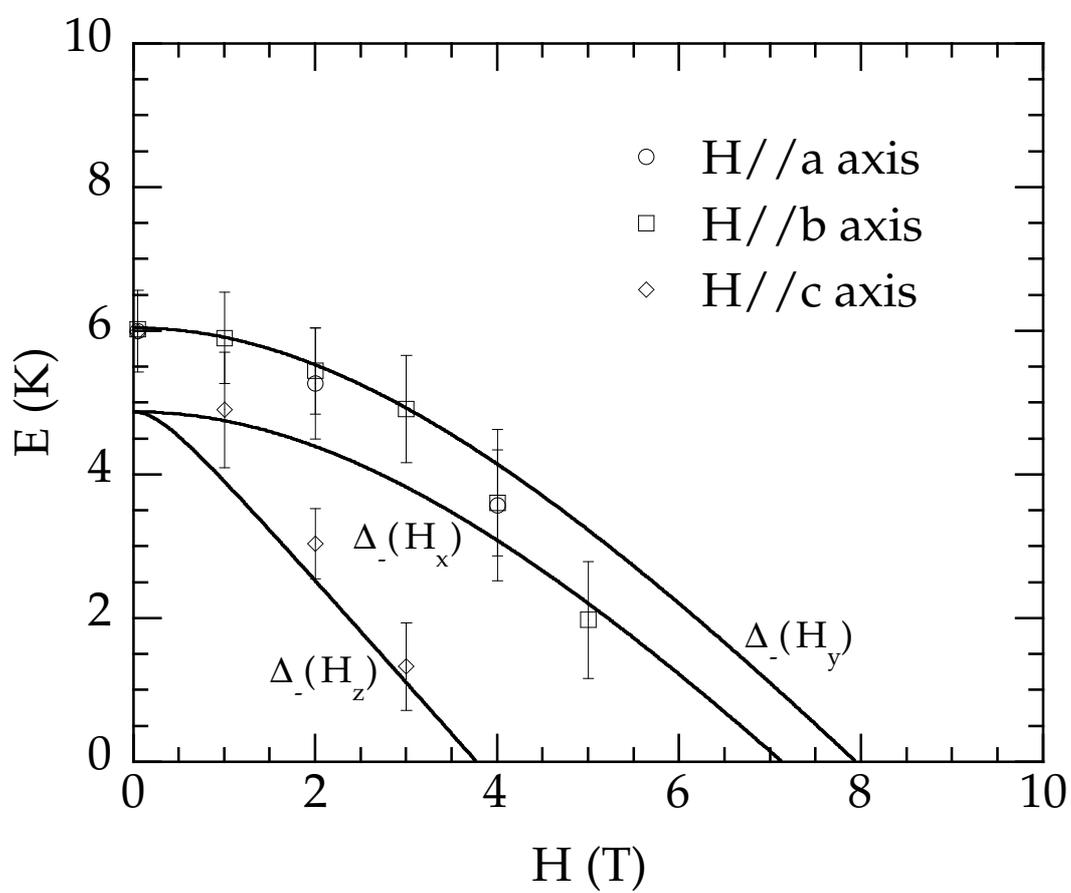
\begin{figure}
\caption{The magnetic field dependence of the energy difference ($\Delta_{-}(\bf H)$) between
the singlet ground state and the lowest level of the excited triplet obtained from the analysis
of $C_{\rm m}$.  The solid curves represent the theoretical result (Ref. 24).}
\label{fig10}
\end{figure}

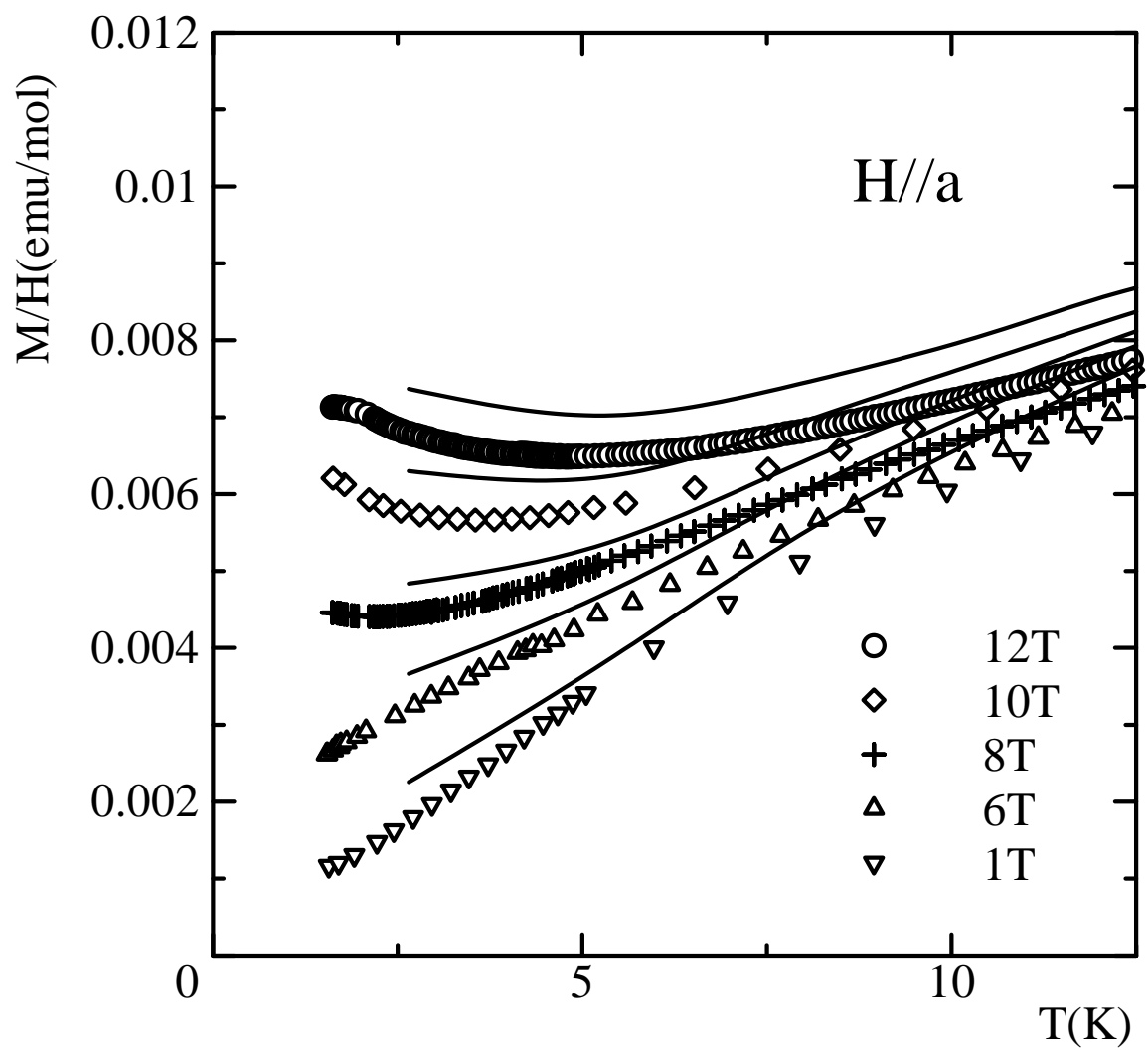
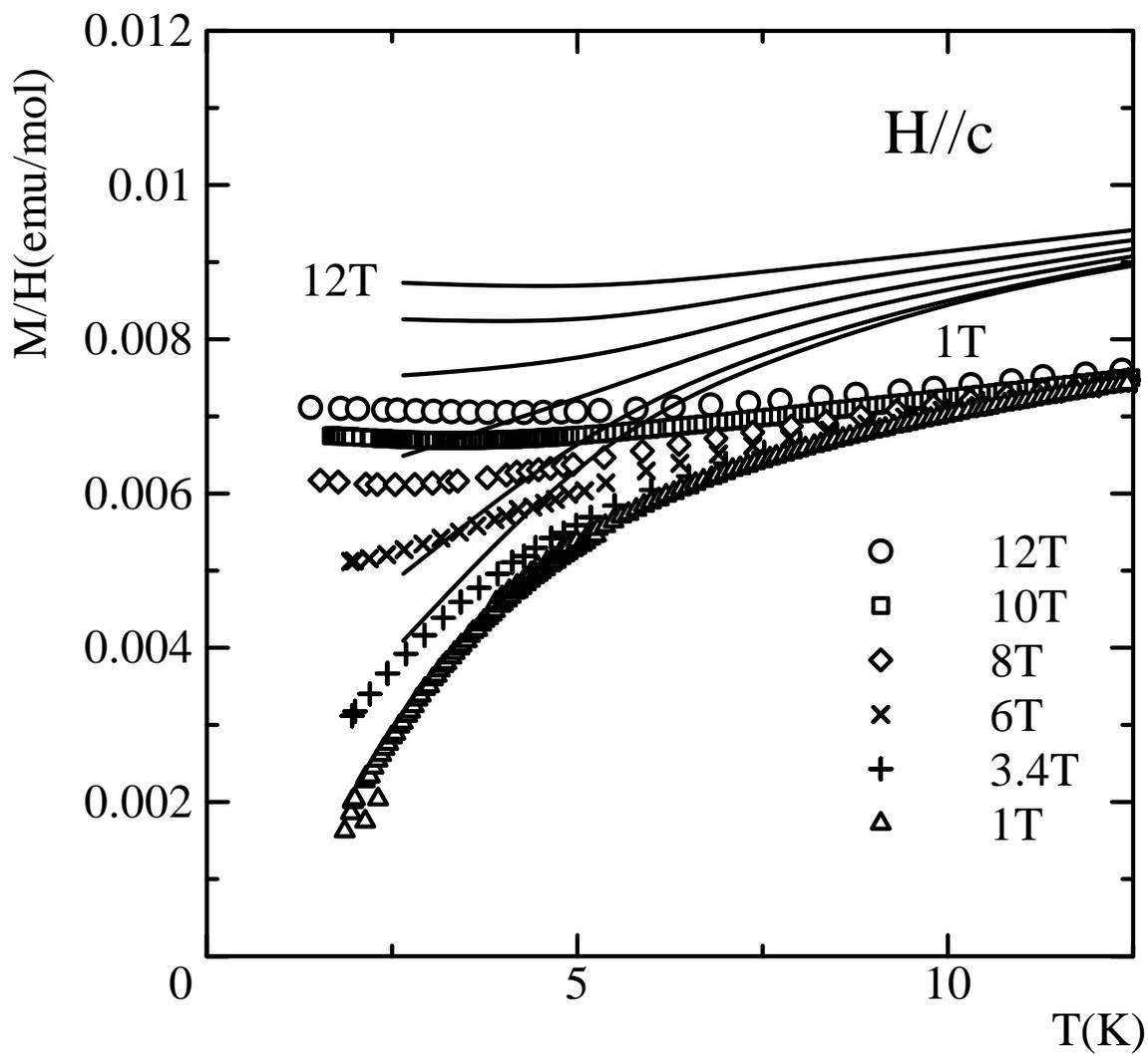
\begin{figure}
\caption{Comparison between theory and experiment on the temperature dependence of
the susceptibility ($M/H$) at finite fields for (a) $H\parallel a$ and (b) $H\parallel c$.
The solid lines are the theoretical ones discussed in the text.}
\label{fig11}
\end{figure}


\begin{references}
\bibitem[*] {ZH00}Also at Graduate School of Science and Engineering, Saitama University, Urawa, 
Saitama 338-8570, Japan
\bibitem{FDH83}F. D. M. Haldane, Phys. Lett. {\bf 93}A, 464 (1983); Phys. Rev. Lett. 
{\bf 50}, 1153 (1983).
\bibitem{IA89}For a review see, I. Affleck, J. Phys. : Condens. Matter {\bf 1}, 3047 (1989).
\bibitem{KK95}For a review see, K. Katsumata, J. Magn. Magn. Mater. {\bf 140-144}, 1595 (1995).
\bibitem{LJ74}For a review see, L. J. de Jongh and A. R. Miedema,
$Experiments$ $on$ $Simple$ $Magnetic$ $Model$ $Systems$ (Taylor \& Francis Ltd.,
London, 1974).
\bibitem{AM82}A. Meyer, A. Gleizes, J. Girerd, M. Verdaguer and O. Kahn, Inorg. Chem. 
{\bf 21}, 1729 (1982).
\bibitem{JPR87}J. P. Renard, M. Verdaguer, L. P. Regnault, W. A. C. Erkelens, J. Rossat-Mignod 
and W. G. Stirling, Europhys. Lett. {\bf 3}, 945 (1987).
\bibitem{OA92}O. Avenel, J. Xu, J. S. Xia, M-F. Xu, B. Andraka, T. Lang, P. L. Moyland, 
W. Ni, P. J. C. Signore, C. M. C. M. van Woerkens, E. D. Adams, G. G. Ihas, M. W. Meisel, 
S. E. Nagler, N. S. Sullivan and Y. Takano, Phys. Rev. B{\bf 46}, 8655 (1992).
\bibitem{DM77}D. Moses, H. Shechter, E. Ehrenfreund and J. Makovsky, J. Phys. C: 
Solid State Phys. {\bf 10}, 433 (1977).
\bibitem{ST90}T. Sakai and M. Takahashi, Phys. Rev. B{\bf 42}, 4537 (1990).
\bibitem{KK89}K. Katsumata, H. Hori, T. Takeuchi, M. Date, A. Yamagishi and J. P. Renard, 
Phys. Rev. Lett. {\bf 63}, 86 (1989).
\bibitem{TK92}T. Kobayashi, Y. Tabuchi, K. Amaya, Y. Ajiro, T. Yosida and M. Date,
J. Phys. Soc. Jpn. {\bf 61}, 1772 (1992).
\bibitem{MC91}M. Chiba, Y. Ajiro, H. Kikuchi, T. Kubo and T. Morimoto, Phys. Rev. 
B{\bf 44}, 2838 (1991).
\bibitem{MH94}P. P. Mitra and B. I. Halperin, Phys. Rev. Lett. {\bf 72}, 912 (1994).
\bibitem{SS94}T. Sakai and H. Shiba, J. Phys. Soc. Jpn. {\bf 63}, 867 (1994).
\bibitem{ZH97}Z. Honda, K. Katsumata, H. Aruga Katori, K. Yamada, T. Ohishi, 
T. Manabe and M. Yamashita, J. Phys.: Condens. Matter {\bf 9}, L83 (1997); 3487 (1997).
\bibitem{HAK98}Z. Honda, H. Asakawa and K. Katsumata, Phys. Rev. Lett. {\bf 81}, 
2566 (1998).
\bibitem{HKH99}Z. Honda, K. Katsumata, M. Hagiwara and M. Tokunaga, Phys. Rev. B {\bf 60}, 
9272 (1999).
\bibitem{MM96}M. Monfort, J. Ribas, X. Solans and M. F. Bardia, Inorg. Chem. {\bf 35}, 
7633 (1996).
\bibitem{AZ00}A. Zheludev, Y. Chen, C. L. Broholm, Z. Honda and K. Katsumata, 
Cond-mat/0003223.
\bibitem{GS93}We thank G. Shirane for suggesting this possibility.
\bibitem{JPA78}See for example, J. P. A. M. Hijmans, K. Kopinga, F. Boersma and 
W. J. M. de Jonge, Phys. Rev. Lett. {\bf 40}, 1108 (1978).
\bibitem{ST91}T. Sakai and M. Takahashi, J. Phys. Soc. Jpn. {\bf 60}, 3615 (1991).
\bibitem{HSH81}I. Harada, K. Sasaki and H. Shiba, Solid State Commun. {\bf 40}, 
29 (1981).
\bibitem{OG93}O. Golinelli, Th. Jolicoeur and R. Lacaze, J. Phys.: Condens. Matter {\bf 5}, 
7847 (1993).
\bibitem{HK00}Z. Honda, K. Katsumata, and M. Hagiwara, unpublished.
\end{references}
\end{document}